\documentstyle[preprint,eqsecnum,aps]{revtex}

\begin{document}

\draft
\preprint{KAIST-CHEP-96/8}

\title{Exact wave functions and geometric phases
of a generalized driven oscillator}
   
\author{Min-Ho Lee\thanks{e-mail:mhlee@chep6.kaist.ac.kr}, 
Hyeong-Chan Kim\thanks{e-mail:leo@chep5.kaist.ac.kr}}
\address{
Department of Physics, Korea Advanced Institute of
Science and Technology, Taejon, 305-701, Korea}
\author{Jeong-Young Ji\thanks{e-mail:jyji@phyb.sun.ac.kr} }
\address{Department of Physics Education,
Seoul National University, Seoul, 151-742, Korea}

\maketitle

\begin{abstract}
The generalized invariant and its eigenstates of a general quadratic
oscillator are found. The Schr\"odinger wave functions for the eigenstates 
are also found in analytically closed forms. The conditions for the existence
of the cyclic initial state (CIS) are studied and the corresponding 
nonadiabatic Berry phase is calculated explicitly.
\end{abstract}



\section{Introduction} \label{sec1}

For a long time, much attention has been concentrated
on the time-dependent harmonic oscillator
both in classical and quantum mechanics.
The quantization of a time-dependent harmonic oscillator is
very important when we treat free quantum fields
in curved spacetime~\cite{Fulling},
quantum statistical properties of radiation~\cite{Louisell73},
coherent states and squeezed states of light~\cite{MeystreS91},
etc.

One of the most powerful methods to find a quantum state of
a time-dependent oscillator is the generalized invariant method 
introduced by Lewis~\cite{lewis} and Lewis and Riesenfeld (LR)~\cite{LR}.
They have derived a simple relation between eigenstates of 
LR  invariant and solutions of the corresponding Schr\"{o}dinger 
equation. The Schr\"{o}dinger wave functions have been studied in
Refs.~\cite{spkim,yeon} and their results are generalized
in Ref.~\cite{jyji2}.

The time-dependent harmonic oscillator with damping and a perturbative
force was considered by Khandekar and Lawande who found the 
exact wave function in terms of an auxiliary function $\rho$ and 
an external force $f(t)$ by using the LR invariant \cite{lawande}.
The exact evolution operator for the harmonic oscillator subject to
an arbitrary force was obtained by Lo~\cite{lo}. He obtained 
the quantum states by solving the differential equation
satisfied by the evolution operator. 
For a generalized harmonic oscillator (GHO) with a constant force, 
the invariant formulation was developed and the geometric phase
was calculated in Ref.~\cite{gao}.
Recently, the authors have calculated the generalized invariant
and its eigenstates for a time-dependent forced harmonic 
oscillator in the Heisenberg picture~\cite{hckim}. In addition,
they compared their quantum states with them of Lo.

In this paper, we extend our previous work~\cite{hckim}
to a general quadratic oscillator.
By adding time-dependent terms proportional to $pq+qp$ and $p$
to the Hamiltonian of Ref.~\cite{hckim} we have the most general 
type of a time-dependent quadratic oscillator.
Hereafter, we call this type of oscillator a generalized driven oscillator (GDO)
to discriminate it from a GHO which has no linear driving terms.
This work is also an extension of Ref.~\cite{gao} in the points
that a driven force is generalized from the constant one 
to a time-dependent one and the $p$ term is not considered in Ref.~\cite{gao}.
For a GDO, we find the LR invariant, its eigenstates 
and the corresponding Schr\"{o}dinger wave functions. 
Analyzing the wave functions, we find the CISs and
calculate the corresponding non-adiabatic Berry phases.

In Sec.~\ref{sec2} we find a generalized invariant for the GDO.
First, we find a generalized invariant for a GHO, where we
represent the invariant in terms of the classical solutions.
For this invariant, we construct the creation and annihilation operators.
Using these operators, we find a generalized invariant for the GDO
employing the method introduced in Ref.~\cite{hckim}.
We also construct the Fock space using the eigenstates of the invariant.
In Sec.~\ref{sec3}
we obtain the exact wave function which satisfies
the time-dependent Schr\"{o}dinger equation.
In Sec. IV we study the condition for the existence of CISs
and calculate non-adiabatic Berry phases.
In Sec.~\ref{sec4}
some wave functions are presented as examples.
The last section is devoted to the discussion. 


\section{Generalized invariant and its eigenstates}\label{sec2}

Consider the Hamiltonian of a GDO:
\begin{equation}
H_T(t) = H(t)
- F(t) q(t)  -  G(t) p(t),
\label{Ham}
\end{equation}
where
\begin{equation}
H(t) =
\frac{1}{2M(t)} p^2(t)
+ \frac{Y(t)}{2} \left[ p(t) q(t) + q(t) p(t) \right]
+  \frac{1}{2} M(t) \omega^2(t) q^2(t)
\label{ham}
\end{equation}
and $M(t), \omega(t)^2$, and $ Y(t)$ are time dependent parameters
that satisfy $\omega^2(t)-Y^2(t)>0$. The $F(t)$ and $G(t)$ terms
describe linear driving mechanism.

Since the set of Hermitian operators,
$\{ \frac{1}{2}p^2(t),\frac{1}{2} q^2(t),
\frac{1}{2} [ p(t) q(t) +  q(t) p(t) ], p(t),q(t),1 \}$,
forms the Lie algebra,
we try to find the following form of the LR invariant:
\begin{equation}
I_T(t) = I(t)
+ g_1 (t)  q(t) + g_2 (t) p(t) + g_3(t)
\label{Inv}
\end{equation}
where
\begin{equation}
I(t) = g_- (t) \frac{p^2(t)}{2}
+ g_0 (t) \frac{ p(t) q(t) + q(t) p(t) }{2}
+ g_+ (t)  \frac{ q^2(t)}{2} .
\label{inv}
\end{equation}

\subsection{Generalized Harmonic oscillator with no linear driving terms}

Before proceeding with the Hamiltonian (\ref{Ham}),
let us consider the Hamiltonian of the GHO~(\ref{ham}).
The invariant for this Hamiltonian will be the form of (\ref{inv}),
and from the invariant equation (in $\hbar=1$ units)
\begin{equation}
\frac{d }{d t} I (t)  =
\frac{ \partial}{ \partial t } I(t) - i [ I(t), H(t)] = 0,
\end{equation}
we obtain a coupled linear system of the first-order differential
equations
\begin{eqnarray}
\nonumber
\dot{g}_- &=& 2 g_- Y - \frac{2}{M} g_0, \\
\dot{g}_0 &=& M \omega^2 g_- - \frac{1}{M} g_+.   \label{eqo} \\
\nonumber
\dot{g}_+ &=& - 2 g_+ Y + 2 M \omega^2 g_0,
\end{eqnarray}
where a dot denotes a time-derivative.
Introducing
\begin{equation}
\bar{g}_- = \alpha^2 g_-,~\bar{g}_0 = g_0,~ \bar{g}_+ = \frac{g_+}{\alpha^{2}},
~\bar{M}(t) = \frac{M(t)}{ \alpha^2},
\label{trf}
\end{equation}
where $ \alpha(t) = e^{ - \int^t  Y d t }$,
it follows from (\ref{eqo}) that
\begin{eqnarray}
\nonumber
\dot{\bar{g}}_- &=&  - \frac{2}{\bar{M}} \bar{g}_0, \\
\dot{\bar{g}}_0  &=&   \bar{M} \omega^2 \bar{g}_-
- \frac{1}{\bar{M}} \bar{g}_+,
\label{eq3} \\
\nonumber
\dot{\bar{g}}_+ &=&  2 \bar{M} \omega^2 \bar{g}_0 .
\end{eqnarray}
Here we note that $\bar{g}_{\pm,0}$ satisfy the same differential equations
as Eq.~(6) in Ref.~\cite{jyji}. Thus $\bar{g}_{\pm,0}$  can be represented
in terms the solutions of the differential equation
$(d/dt) [\bar{M} \dot{\bar f}] + \bar{M} \omega^2 \bar f = 0$.
Further, we have the relations
$\bar{f}_1  = \alpha f_1, \bar{f}_2 = \alpha f_2$,
where $f_{1,2}(t)$ are the two independent solutions of
the equation of motion for $H(t)$:
\begin{equation}
M \ddot{f}_i + \dot{M} \dot{f}_i + ( M \omega^2 - M Y^2
- \dot{M} Y - M \dot{Y} ) f_i = 0.
\label{eom}
\end{equation}
Therefore the most general solutions of Eq.~(\ref{eq3}) are given by
\begin{eqnarray}
\nonumber
g_-(t) &=& c_1 f_1^2 (t) + c_2 f_1 (t) f_2 (t) + c_3 f_2^2 (t), \\
\nonumber
g_0 (t) &=& - M(t)  \left( c_1 f_1(t)  \dot{f_1}(t)
   + \frac{c_2}{2} \left[ \dot{f_1}(t)  f_2 (t) +
  f_1(t)  \dot{f_2} (t) \right] + c_3 f_2(t)
  \dot{f_2} (t) \right) + M(t) Y(t) g_-(t),      \label{solo} \\
g_+(t)  &=& M^2(t)  \left( c_1 \dot{f_1}^2 (t) + c_2 \dot{f_1}(t)
\dot{f_2} (t) + c_3 \dot{f_2}^2 (t) \right) + M(t) Y^2(t) g_-(t) +
2 Y(t) g_0(t),
\end{eqnarray}
where $c_1,c_2$ and $c_3$ are arbitrary constants.

Now the invariant (\ref{inv}) can be written as
\begin{equation}
I(t) = \omega_I \left(
  b^\dagger (t) b(t) + \frac{1}{2}   \right),
\label{Ib}
\end{equation}
introducing
\begin{eqnarray}
\nonumber
b(t) &=&
\left( \sqrt{\frac{\omega_I}{2 g_-(t)} }+ i \sqrt{\frac{1}{2\omega_I
    g_-(t)}} g_0(t)
   \right) q(t) + i\sqrt{\frac{g_-(t)}{2 \omega_I}} p(t), \\
b^{\dagger}(t) &=&
\left( \sqrt{\frac{\omega_I}{2 g_-(t)} }- i \sqrt{\frac{1}{2\omega_I
    g_-(t)}} g_0(t)
   \right) q(t) - i\sqrt{\frac{g_-(t)}{2 \omega_I}} p(t)
   \label{acop}
\end{eqnarray}
with $\omega_I^2  =  g_+ (t) g_-(t) - g^2_0 (t)$.
From the equation of motion for $b(t)$:
\begin{equation}
\frac{d}{dt} b(t) = \frac{\partial}{\partial t} b(t)
- i \left[b(t), H(t) \right] \\
= -i \frac{\omega_I}{M(t) g_-(t)} b(t),
\label{bY}
\end{equation}
we have
\begin{equation}
b(t) = b e^{-i \Theta(t)},
\label{bsol}
\end{equation}
where
\begin{equation}
\Theta(t) = \int^t_{t_0} dt' \frac{\omega_I}{M(t') g_-(t')} .
\end{equation}
Then it is clear from (\ref{Ib}) and (\ref{bsol})
that $I(t)$ is an invariant for $H(t)$.


\subsection{General quadratic Hamiltonian with linear driving}

Now we find the invariant (\ref{Inv}) for the total Hamiltonian (\ref{Ham}).
Inserting the (\ref{Ham}) and (\ref{Inv}) into
\begin{equation}
\frac{d }{d t} I_T(t)  =
\frac{ \partial}{ \partial t } I_T(t) - i [ I_T(t), H_T (t)] = 0,
\end{equation}
we get the linear coupled differential equations (\ref{eqo})
and additionally we have for $g_{1,2,3} (t)$:
\begin{eqnarray}
\nonumber
\dot{g}_1 &=&  - Y g_1 + M \omega^2 g_2  + ( G g_+ - F g_0 ), \\
\dot{g}_2 &=& - \frac{1}{M} g_1  + Y g_2 - ( F g_- - G g_0 ),
\label{eqt} \\
\nonumber
\dot{g}_3 &=& G g_1 - F g_2 .
\end{eqnarray}
However these equations are too difficult to get the solutions,
and we get around this difficulty using the method introduced
in Ref.~\cite{hckim}.

For the presence of the linear driving terms
we search for the invariant of the form
\begin{equation}
I_T(t)  =
 \omega_I \left( B^\dagger(t) B(t) + \frac{1}{2} \right),
 \label{IB}
\end{equation}
where
\begin{equation}
B(t) = b(t) + \beta(t).
\label{Bb}
\end{equation}
It follows from
\begin{eqnarray}
\frac{d}{d t} b(t) &=& \frac{\partial}{\partial t} b(t)
      - i \left[b(t), H_T(t) \right] \\
  &=& -i \frac{\omega_I}{M(t) g_-(t)} b(t)  + W(t)  \nonumber,
\end{eqnarray}
with
\begin{equation}
W(t)  = - \left( \sqrt{\frac{\omega_I}{2 g_-(t)} }
   + i \sqrt{\frac{1}{2\omega_I
    g_-(t)}}  g_0 (t)
   \right)  G(t)
   + i \sqrt{\frac{g_-(t)}{2 \omega_I}} F(t)
\label{Wt}
\end{equation}
that
\begin{equation}
\frac{d}{d t} B(t)
  = -i \frac{\omega_I}{M(t) g_-(t)} B(t),  \label{basic}
\end{equation}
provided that $ \beta(t)$ satisfies the differential equation
\begin{equation}
\frac{d}{dt} \beta(t)  + i \frac{\omega_I}{M(t) g_-(t) } \beta(t)
  = - W(t).
\label{betaEq}
\end{equation}
The solutions of (\ref{basic}) and (\ref{betaEq}) are
easily found:
\begin{equation}
B(t)  = e^{ -i \Theta (t) }B(t_0),~~
B^\dagger (t)  = e^{  i \Theta (t) }B^\dagger (t_0),
\end{equation}
and
\begin{equation}
\beta(t) =  e^{ - i \Theta (t) } \left(
   \beta_0   - \int^t W(t')  e^{i \Theta (t') } dt'  \right),
\label{beta}
\end{equation}
where $\beta_0$ is an integration constant.

Now it is trivial to find $g_i(t)~(i=1,2,3)$,
by substituting (\ref{acop}) into (\ref{IB})
using (\ref{Bb}) we get
\begin{eqnarray}
\nonumber
I_T(t)   &=& I (t)
   + \omega_I \beta_R \sqrt{ \frac{ 2 \omega_I}{g_-(t)}} q(t)
   + \omega_I \beta_I \sqrt{\frac{ 2 g_- (t)}{\omega_I}}
   \left(
    p(t) + \frac{ g_0 (t)}{g_-(t)} q(t)  \right)
    \\
  & & ~  + ~ \omega_I ( \beta_R^2 + \beta_I^2 ),
\label{inv2}
\end{eqnarray}
where $\beta_R ~(\beta_I) $ is the real (imaginary) part of
$\beta$.
By comparing (\ref{Inv}) and (\ref{inv2}),
we can easily read that
\begin{eqnarray}
g_1 (t) &=& 
    \omega_I \beta_R \sqrt{ \frac{ 2 \omega_I}{g_-(t)}} 
   + \omega_I \beta_I \sqrt{\frac{2 g_-(t)}{\omega_I} }
           \frac{ g_0 (t)}{g_-(t)},   \\
g_2(t) &=&  \omega_I \beta_I \sqrt{\frac{ 2 g_-(t)}{\omega_I} }, \\ 
g_3(t) &=&    \omega_I ( \beta_R^2 + \beta_I^2 ).
\end{eqnarray}
Thus we have found the generalized invariant 
-- (\ref{Inv}) or (\ref{IB}) --
for the GDO~(\ref{Ham}).
                                        
As in the usual case,
we construct the Fock space as eigenstates of $I_T (t)$,
which are given by
\begin{eqnarray}
\left| n, t\right> = \frac{{B^{\dagger}}^n(t)}{\sqrt{n !}}
                                \left|0, t\right>,
\end{eqnarray}
where
\begin{equation}
B(t) \left| 0, t \right> = 0.
\end{equation}
The eigenvalue equation of $I_T(t)$ can be written as
\begin{equation}
I_T(t) \left| n,t \right> =  \omega_I
\left(n + \frac{1}{2} \right) \left| n, t \right>,~
n = 0,1,2,3,....
\end{equation}

\section{Wave Function of  the Schr\"{o}dinger Equation}
\label{sec3}

Following LR~\cite{LR} the wave function of the 
Schr\"{o}dinger equation 
\begin{equation}
i   \frac{\partial}{\partial t} \left| \psi_n ( t )\right>_S =
H_T (t) \left| \psi_n ( t) \right>_S
\end{equation}
can be written by
\begin{equation}
\left|\psi_n(t) \right>_S
= e^{i \alpha_n (t) }  \left| n,t \right>,
\label{SchKet}
\end{equation}
where the phase functions $\alpha_n(t)$ are found from the equation
\begin{equation}
\frac{d}{dt} \alpha_n (t) = 
\left< n,t \right| \left (i \frac{ \partial}{\partial t} 
    - H_T(t) \right)  \left| n,t \right>.
    \label{phase}
\end{equation}

\subsection{ The coordinate representation of the eigenstates of the invariant}  
Let us define new coordinate $\bar{Q}$ and momentum
$\bar{P}$ by taking the following time-dependent successive unitary 
transformations:
\begin{eqnarray}
Q(t) &=&  U_1^\dagger (t) q(t) U_1 (t) = q(t), \\
P(t) &=&  U_1^\dagger (t) p(t) U_1 (t) = 
           p(t) + \frac{g_0(t)}{g_-(t)} q(t), \\
\tilde{Q}(t) &=& U_2^\dagger(t) Q (t) U_2 (t) = 
             Q(t) + \Delta_q(t), \\
\tilde{P}(t) &=& U_2^\dagger(t) P (t) U_2(t) = 
             P(t) + \Delta_p(t),\\ 
\bar{Q}(t) &=& U_3^\dagger (t)  \tilde{Q}(t) U_3 (t) 
         = \frac{1}{\sqrt{g_-(t)}} \tilde{Q}(t), \\
\bar{P}(t) &=& U_3^\dagger (t)  \tilde{P}(t) U_3 (t) 
         = \sqrt{g_-(t)} \tilde{P}(t), 
\end{eqnarray}
where
\begin{eqnarray}
U_1(t) &=&  \exp{ \left(
             i \frac{ g_0 (t) }{g_-(t)} q^2 \right)},  
             \label{u1}  \\
U_2(t) &=& \exp{\left(  - i  \Delta_q (t) P \right)}~ 
            \exp{ \left(i  \Delta_p(t)  Q \right)}, 
            \label{u2}   \\
U_3 (t)  &=&  \exp{ \left(
              \frac{i}{4} (\tilde{P} \tilde{Q} + \tilde{Q} \tilde{P} ) 
           \ln g_-(t) \right)  },
           \label{u3}
\end{eqnarray}
with
\begin{equation}
\Delta_q(t)  = \sqrt{ \frac{ 2 g_-(t)}{\omega_I }} \beta_R(t), ~~~
\Delta_p (t) = \sqrt{ \frac{ 2 \omega_I}{g_-(t)}} \beta_I(t).
\end{equation}
Then   the LR invariant  can be rewritten as a simple form
\begin{equation}
I_T(t)  = \frac{1}{2}   \bar{P}^2(t) + \frac{1}{2} \omega^2_I
           \bar{Q}(t)^2.
\end{equation}
The eigenfunction of $I_T(t)$ in $\bar{Q}$ coordinate is well known as
\begin{equation}
\left<  \bar{Q},t | n \right> =
\frac{1}{\sqrt{2^n n! } } \left( 
\frac{\omega_I}{\pi} \right)^{1/4} e^{ 
        -  \frac{\omega_I \bar{Q}^2 }{2}} H_n \left(
           \sqrt{\omega_I} \bar{Q} \right),
\end{equation}
where $H_n $ is a Hermite polynomial
and  which is normalized to be 
 $\left< n,t| n',t \right> = \delta_{nn'}$.

Now we represent the wave function in terms of
original coordinate variable
\begin{equation}
\left<  q  | n, t \right> \equiv  \varphi_n (q,t) = \left<  q,t | n \right>.
\end{equation}
It follow from Eqs. (\ref{u1}), (\ref{u2}), and (\ref{u3}) 
that
\begin{eqnarray}
\left<  q,t \right|
\nonumber
  &=&      e^{- i \frac{g_0}{ g_-} q^2 }
  \left<  Q,t \right|   \\
\nonumber
  &=&    e^{- i \frac{g_0}{ g_-} q^2 
  - i Q \Delta_p  } \left<  \tilde{Q},t \right|  \\
\nonumber
  &=&    e^{- i \frac{g_0}{ g_-} q^2 
  - i Q \Delta_p  }  \left( g_- \right)^{-1/4} 
  \left<  \bar{Q},t \right|  \\
  &=&   \left( \frac{1}{g_- } \right)^{1/4} 
  e^{ - i \frac{g_0}{ g_-} q^2 
  - i  \Delta_p q  } 
  \left<  \bar{Q},t \right|,
\end{eqnarray}
and
therefore the coordinate representation of the invariant eigenstate
is  found to be 
\begin{equation}
\varphi_n (q,t) \equiv \left< q| n,t \right> =
\frac{1}{\sqrt{2^n n!}} 
\left( \frac{\omega_I}{\pi g_-}  \right)^{ \frac{1}{4} }
e^{ -  i \frac{g_0}{2 g_-} q^2  - i \Delta_p  q 
 - \frac{\omega_I}{2 g_- } ( q + \Delta_q)^2 } 
 H_n \left( \sqrt{\frac{\omega_I}{g_-}} \left( q  + \Delta_q \right)  
 \right).
 \label{wave}
\end{equation}
We  can show that Eq.~(\ref{wave})  satisfies the equation 
$I_T(t) \varphi_n(q,t) = \omega_I (n + \frac{1}{2})
\varphi_n (q,t)$ by direct differentiations.
It should be noted that the coordinate representation of the eigenstate
is not unique since the time-dependent phase is arbitrary. 
Consequently, we should find $\alpha_n (t)$ in (\ref{SchKet}) 
by solving Eq.~(\ref{phase}) to find the wave function 
satisfying the Schr\"odinger equation.

\subsection{Schr\"odinger Wave Function}
As in Ref.~\cite{LR}, we find the phase $\alpha_n (t)$ in Eq.~(\ref{SchKet})
using the following relations:
\begin{equation}
\left< n,t \right| \frac{\partial}{\partial t} \left| n, t \right>
= \left< n- 1,t \right| \frac{\partial}{\partial t} 
\left| n - 1, t \right>
  + \frac{1}{ \sqrt{n} }
  ~\left< n,t \right| \frac{\partial B^\dagger}{\partial t} 
\left| n, t \right> ,
\end{equation}
\begin{equation}
\frac{\partial}{\partial t} B^\dagger (t) = 
i \frac{\omega_I}{ M(t) g_-(t) }  B^\dagger (t)  + i \left[
B^\dagger (t),  H_T(t) \right]
\end{equation}
and
\begin{eqnarray}
\nonumber
H_T(t) &=& H_Q(t) 
- B(t) \left(
h_-(t)\beta(t) + \frac{1}{2} h_0(t) \beta^\dagger (t) + i W^\dagger
\right)  
\label{HT:B} \\
\nonumber
&-& B^\dagger(t) \left(
h_+(t)\beta_+(t) + \frac{1}{2} h_0(t) \beta (t) - i W(t)
\right)  \\
&+&\left( i W^\dagger(t) \beta(t) - i W \beta^\dagger(t) \right)
+ \frac{1}{2}
\left( h_-(t) \beta^2(t) + h_0 \beta(t) \beta^\dagger (t) + h_+(t) 
{\beta^\dagger}^2(t) \right),
\end{eqnarray}
where
\begin{equation}
H_Q(t) = h_+(t)\frac{B^{\dagger(t) 2}}{2} 
       + h_0(t) \frac{B(t) B^\dagger(t)
       +B^\dagger(t)B(t)}{4} + h_-(t) \frac{B(t)^2}{2}
\end{equation}
and 
\begin{eqnarray}
\nonumber
h_0(t) &=& \frac{1}{M(t) g_-(t) \omega_I}
\left[
g_0^2(t) + M^2(t)\omega^2(t) g_-^2(t) + \omega_I^2 - 2 M(t)Y(t)g_0(t)
g_-(t)
\right] \\
\nonumber
h_\pm(t) &=& \frac{1}{2 M(t) g_-(t)\omega_I}
\left[
g_0^2(t) + M^2(t)\omega^2(t) g_-^2(t) - \omega_I^2 \mp
2 i g_0(t) \omega_I  \right. \\
& &  \left.  - 2 M(t)Y(t)g_0(t) g_-(t)
\pm i 2 M(t) Y(t) g_-(t) \omega_I
\right].
\end{eqnarray}
After a little algebra, we obtain the phase
\begin{eqnarray}
\nonumber
\alpha_n (t) &=& - ( n + \frac{1}{2} )  \int^t dt^{'} 
\frac{\omega_I}{ M(t') g_-(t')}  \\
&+&   \int^t dt' \left[  
     \left(  \beta_R^2(t')  - \beta_I^2 (t')\right) 
  \frac{ \omega_I}{M(t') g_-(t')}
     - \sqrt{  \frac{2 \omega_I }{g_-(t')} } G(t') \beta_I(t') \right].
\end{eqnarray}
Therefore, we find the exact solution of the Schr\"{o}dinger equation:
\begin{eqnarray}
\nonumber
\psi_n (q,t)  &=&    e^{ -  i
  ( n + \frac{1}{2} )   \int \frac{\omega_I}{
    M g_-} d t  + i \int  dt \left[
  \left(  {\beta_R}^2  - {\beta_I}^2 \right) \frac{ \omega_I}{ M g_-}
     - \sqrt{  \frac{2 \omega_I }{g_-} } G \beta_I \right] 
     }      \\
& & ~~\times \frac{1}{\sqrt{2^n n!}} 
\left( \frac{\omega_I}{\pi g_-}  \right)^{ \frac{1}{4} }
e^{ -  i \frac{g_0}{2 g_-} q^2  - i \Delta_p  q 
 - \frac{\omega_I}{2 g_- } ( q + \Delta_q)^2 } 
 H_n \left( \sqrt{\frac{\omega_I}{g_-}} \left( q  + \Delta_q \right) 
 \right).
\label{WF}
\end{eqnarray}
This wave function becomes the one of Ref.~\cite{jyji} 
when $F(t) = 0 = G(t)$ and $ Y(t) = 0$.
\section{Existence of cyclic initial states and nonadiabatic Berry phases}
In this section we examine carefully the wave function (\ref{WF})
to find CISs and the corresponding non-adiabatic Berry phases.  
Suppose that all parameters $M(t)$, $Y(t)$, $\omega(t)$, $F(t)$ and $G(t)$ 
are periodic function of $t$ so that the Hamiltonian is cyclic with a period $\tau$.
The definition of CIS gives $\psi(t_0 + \tau) = e^{i \chi} \psi(t_0)$, 
and the Berry phase $\Gamma$ is given by
$\Gamma = \chi - \delta$,
where
$\delta = -  \int_0^{\tau} \langle \psi(t) | H(t) | \psi(t) \rangle dt$.

Inspecting closely the form of the wave function (\ref{WF}),
we find that the necessary and sufficient condition for the existence 
of CISs is the existence of periodic $g_-(t)$ and $\beta(t)$.
The periodic property of $g_-(t)$ is related to the solution of 
Eq.~(\ref{eom}). The conditions for the periodic solutions of classical
equation of motion were well studied in Ref.~\cite{jyji}.
Therefore we investigate only the condition for the periodic $\beta(t)$.
Hereafter we assume that $g_-(t)$ is a $\tau$-periodic function.

To study the periodicity of $\beta(t)$, from (\ref{beta}) we write
\begin{eqnarray}
\beta(t+\tau) &=& e^{ -i \Theta(t+\tau) }
  \left( \beta_0 - \int^{t+\tau} W(t') e^{i \Theta(t')} dt' \right)
\nonumber \\
&=& e^{-i \theta_0} \left[
\beta(t) - \sigma_0 e^{ - i [\Theta(t) } \right] ,
\label{pe_beta}
\end{eqnarray}
where we used Eq.~(\ref{beta}) again. Here $\theta_0$ and $\sigma_0$ are
defined as 
\begin{equation}
\theta_0 = \int_t^{t+\tau} dt'\frac{\omega_I}{M(t') g_-(t') }
\label{th0}
\end{equation}
and
\begin{equation}
\sigma_0 = \int_t^{t+\tau} W(t') e^{ i \Theta(t')} dt' .
\label{si0}
\end{equation}
As seen from the periodicity of the integrand of (\ref{th0}), $\theta_0$ is constant.
Then, if $e^{-i \theta_0 }=1$, the periodicity of $F(t)$ and $G(t)$ guarantees 
the periodicity of the integrand of (\ref{si0})[see (\ref{Wt})], 
and hence $\sigma_0$ is also constant. 
Then it is clear from (\ref{pe_beta}) that the conditions for periodic $\beta(t)$ are 
({\it i}) $\theta_0 = 2 \pi n (n={\rm integers})$ ({\it ii}) $\sigma_0 = 0$.
When these conditions are fulfilled, all the eigenstates of the generalized 
invariant are CISs with
\begin{equation}
\chi_n = \alpha_n (t+\tau) - \alpha_n (t) 
= \alpha_n (\tau) - \alpha_n (0) 
\end{equation}
and the nonadiabatic Berry phase is obtained by 
\begin{equation}
\Gamma_n = \chi_n + \int_{0}^{\tau} \langle \psi_n(t) |
H(t) | \psi_n (t) \rangle dt.
\end{equation}
Using (\ref{HT:B}), we get
\begin{eqnarray}
\Gamma_n (\tau) &=&
\left( n + \frac{1}{2} \right)
\int_{0}^{\tau} dt \left[
\frac{g_0^2(t)}{M(t) g_-(t) \omega_I} - 
\frac{Y (t)g_0(t)}{\omega_0} \right] 
\nonumber \\
&& + \int_{0}^{\tau} dt 
\left[ \frac{h_0(t)}{2} \beta(t) \beta^*(t) 
+ \xi_R + \zeta_R \right]
\label{bp}
\end{eqnarray}
where $\xi_R$ ($\zeta_R$) is the real part of $\xi$ ($\zeta$)
with
\begin{equation}
\xi =
2 i \left( W^* (t) + \sqrt{ \frac{\omega_I}{2 g_-(t)} } \right) 
\beta (t) ,
\end{equation}
\begin{equation}
\zeta = \left( h_-(t) + \frac{\omega_I}{M(t) g_-(t)} \right) \beta^2(t).
\end{equation}
Here we have taken partial integration using Eq.~(\ref{eqo})
to get the first term of (\ref{bp}) as in Ref.~\cite{jyji}.
When $F(t)$, $G(t)$ and $Y(t)$ vanish, this Berry phase
reduces to the one of Ref.~\cite{jyji}. Note that the minus sign 
of Eq.~(4.8) [Eq.~(5.2)] of Ref.~\cite{jyji} is an error. This error can be
easily checked from another expression for the Berry phase Eq.~(4.7) there.

\section{Examples} \label{sec4}
In this  section  we consider  some physically interesting examples
and write the explicit forms of their wave functions and 
study on their Berry phases.

{\it Example A.} $M(t) = m$, $\omega(t) = \omega_0$, $F(t) = F_0$,
and $Y=0=G$:
The two independent solutions of (\ref{eom}) are chosen to be 
$f_1(t) = e^{i \omega_0 t} $ and $ f_2 (t) = e^{ - i \omega_0 t }$.
We can set $g_-(t) = 1/m$, $\omega_I = \omega_0$ by setting
$c_1=0=c_3,~c_2=1/m$ in (\ref{solo}).
Further, it follows from (\ref{betaEq}) that
\begin{equation}
\frac{d}{dt} \beta (t) + i \omega_0 \beta(t) = 
- i \frac{F_0}{\sqrt{ 2 m \omega_0}}.
\end{equation}
and its general solution is given by (\ref{beta}) as
\begin{equation}
\beta = e^{- i \omega_0 t} \left( \beta_0 + 
\frac{F}{\omega_0 \sqrt{2 m \omega_0} } \right) -
 \frac{F}{\omega_0 \sqrt{2 m \omega_0} }
\end{equation}
It is straightforward to find the wave function from (\ref{WF}),
however we confine ourselves to the special case, where
$ \beta (t) = - \frac{F_0}{ \omega_0 \sqrt{ 2 m \omega_0} }$.
Then the wave function is given by
\begin{eqnarray}
\psi_n (q,t) &=&
e^{- i (n + \frac{1}{2} ) 
\omega_0 t + i \frac{F_0^2}{2 m \omega_0^2} t }
\frac{1}{\sqrt{2^n n!}} 
\left( \frac{ \omega_0 m }{\pi} \right)^{1/4}
\nonumber \\
&& \times e^{- \frac{m \omega_0}{2} ( q  - \frac{F_0}{m \omega_0^2} )^2 }
H_n \left(
   \sqrt{m \omega_0} \left( q - F_0 / m \omega_0^2 \right) 
   \right),
\end{eqnarray}
which shows clearly that the wave function of the oscillator 
is shifted by the amount of $F_0/m \omega_0^2$. 
We can always take an arbitrary period $\tau$, then
$\theta_0 = \omega_0 \tau = \frac{2 \pi n}{\tau}$, i.e. 
$\tau = \frac{2 \pi n}{\omega_0}$. Further, $\sigma_0 = 0$.
Thus all  the eigenfunctions of the invariant of this system are
CISs, but all the Berry phases vanish as expected from
the time-independency of the system.

{\it Example B.} The Caldirola-Kanai oscillator \cite{cal,kanai}: 
The Hamiltonian of the Caldirola-Kanai oscillator with external 
force is given by
\begin{equation}
H(t) = \frac{p^2}{2 m e^{2 \gamma t} } +
 e^{ 2 \gamma t}
\left( \frac{m \omega^2 }{2} q^2 +  f(t) q \right),
\label{ckham}
\end{equation}
which describes, at classical level,  a oscillator with a 
time-dependent frequency and a velocity dependent damping term.
The equation of motion without the external force is given by
\begin{equation}
\ddot{q} (t) +  2 \gamma  \dot{q} (t)  + \omega^2  q(t) = 0
\end{equation}
whose solutions   are 
\begin{equation}
f_1(t) = e^{ - \gamma t + i \Omega t}, ~~
f_2(t) = e^{ - \gamma t - i \Omega t},
\end{equation}
where $\Omega = \sqrt{\omega^2 -\gamma^2}$.
By setting $c_1= 0 = c_3 $ and $c_2=1/m$ we get
\begin{equation}
g_- (t) = \frac{1}{m} e^{- 2 \gamma t},~~
g_0(t)  = \gamma,~~
g_+ (t ) = m \omega^2 e^{2 \gamma t }.
\end{equation}
Then, the wave function is given  by
\begin{eqnarray}
\nonumber
\psi_n (q,t) &= &
 e^{ - i ( n + \frac{1}{2} ) \Omega t  +
 i \int^t dt^{'} \left[
   \beta_R^2  - \beta_I^2 \right] \Omega  }~
   \frac{1}{\sqrt{2^n n! }}
   \left( \frac{\Omega m e^{2 \gamma t}}{\pi}  \right)^{1/4}
    \\
\nonumber
& & \times \exp\left(
- i \frac{m \gamma }{2} e^{2 \gamma t } q^2
- i \sqrt{2 m \Omega} e^{ \gamma t} \beta_I q
- \frac{m \Omega }{2} e^{2 \gamma t}
\left( q + \sqrt{\frac{2}{m \Omega}} 
e^{ - \gamma t} \beta_R  \right)^2 \right)
\\
& & \times 
H_n \left( 
        \sqrt{m \Omega} e^{ \gamma t } \left( q +
\sqrt{\frac{2}{m \Omega}} e^{ - \gamma t} \beta_R  \right)
      \right), \label{exwave2}
\end{eqnarray}
where the function $\beta (t) $  is the solution of the following
equation
\begin{equation}
\frac{d}{d t} \beta(t) + i \Omega \beta(t)  =  i 
\frac{f(t) }{\sqrt{2 m \Omega }}
e^{\gamma t},
\end{equation}
and its solution is given by (\ref{beta}).
Unless $\gamma = 0$, the function $g_-(t)$ is not periodic and
this system has no CIS.

{\it Example C.} Special case of {\it Example B} with $\gamma =0$, 
$f(t) = - F_0 \sin ( \omega_{e} t )$, with $( \omega_{e} \neq 0)$.
In this case we have
\begin{equation}
g_-(t) = \frac{1}{m},~
g_0(t) = 0,~
g_+(t) = m\omega^2,
\end{equation}
and 
\begin{equation}
\frac{d}{d t} \beta(t) + i \omega \beta(t)  =  - i 
\frac{F_0 \sin ( \omega_{e} t ) }{\sqrt{2 m \omega }}. 
\end{equation}
The solution $\beta(t)$ is given by
\begin{equation}
\beta(t) =  \beta_0 e^{-i \omega t}
+ i \frac{F_0}{ 2 \sqrt{2 m \omega}} 
\left(
   \frac{e^{i \omega_{e} t } -e^{-i \omega t}}{(\omega + \omega_{e})}
  - \frac{e^{-i \omega_{e} t }-e^{-i \omega t}}{(\omega - \omega_{e})}
\right),
\end{equation}
for $\omega \neq \omega_{e}$. For $\beta(t)$ to be periodic,
it is sufficient that $\omega / \omega_{e}$ is a rational number,
i.e. $\omega / \omega_{e} = r / r_e $, where $r$ and $r_e$ are
positive integers not commensurate. Then the system
has the CISs with a period 
$\tau = 2 \pi r/ \omega = 2 \pi r_e / \omega_e $. 
One note that the period of this system is $\tau_e = 2 \pi / \omega_e$,
while the CISs are $r_e \tau_e$-periodic. 
Using $h_0 = 2 \omega$ and $h_- = 0$, from (\ref{bp}), 
the Berry phase is given by adding the following three terms:
\begin{equation}
\int_0^\tau h_0 (t) |\beta (t)|^2 dt = 
2 \pi r 
\left[ |\beta_0^2| + 
\frac{F_0 (\beta_0 - \beta_0^*) \omega_e }
{i \sqrt{2 m \omega} (\omega^2 - \omega_e^2)}
+ \frac{F_0^2}{2 m \omega}  
\left(
\frac{\omega^2 + \omega_e^2}{(\omega^2-\omega_e^2)^2}
- \frac{1}{2(\omega^2 - \omega_e^2)}  
\right) \right],
\end{equation}
\begin{equation}
\int_0^\tau \xi_R dt = - 2 \pi r \frac{F_0^2}{2 m \omega}
\frac{1}{\omega^2 - \omega_e^2}  ,
\end{equation}
and
\begin{equation}
\int_0^\tau \zeta_R dt = 2 \pi r \frac{F_0^2}{4 m \omega}
\frac{1}{\omega^2 - \omega_e^2} , 
\end{equation}
as
\begin{equation}
\Gamma_n =
2 \pi r 
\left[ |\beta_0^2| + 
\frac{F_0 (\beta_0 - \beta_0^*) \omega_e}
{i \sqrt{2 m \omega} (\omega^2 - \omega_e^2)}
+ \frac{F_0^2}{2 m \omega}  
\left(
\frac{\omega^2 + \omega_e^2}{(\omega^2-\omega_e^2)^2}
- \frac{1}{\omega^2 - \omega_e^2}  
\right) \right].
\end{equation}
It should be noted that, for this example of a forced harmonic oscillator, 
the Berry phase of the eigenstate $\left| n, t \right>$
is independent of quantum number $n$ 
[in Eq.~(\ref{bp}) the first integral term which includes $n$, vanishes].
 
{\it Example D.} The damped pulsating oscillator\cite{lo}:
\begin{equation}
M(t) = m_0 e^{2 ( \gamma  t + \mu \sin \nu t )},~~
\omega^2(t) = \Omega^2  + \frac{1}{ \sqrt{M(t) }}
\frac{d}{dt}  \sqrt{M(t)} .
\end{equation}
The motions of quantum operators are analyzed in Ref.~\cite{hckim}.  
Now we find the wave function using the classical solutions and 
the auxiliary functions used in Ref.~\cite{hckim}: 
\begin{eqnarray}
\nonumber
\psi_n (q,t) &=&
e^{ - i ( n + \frac{1}{2} ) \Omega t + i \int^t dt^{'} 
 \Omega \left[ 
    \beta_R^2 - \beta_I^2 \right]
    }~
    \frac{1}{\sqrt{2^nn!}}
    \left( \frac{ m_0 \Omega}{\pi} \right)^{1/4}
    e^{ \frac{1}{2}  \gamma t  + \frac{1}{2} \mu \sin \nu t }
    \\
\nonumber
& & \times \exp\left( - i \frac{1}{4} \frac{d }{dt} m(t)  q^2   
- \Delta_p q - 
\frac{1}{2} m_0 \Omega e^{ 2 \gamma t + 2 \mu \sin \nu t }
\left(
  q + \Delta_q \right)^2  \right)  \\ 
& & \times H_n \left(
  \sqrt{m_0 \Omega } e^{\gamma t  + \mu \sin \nu t } \left(
  q + \Delta_q \right)  \right),
\end{eqnarray}
where
\begin{equation}
\Delta_q  = \sqrt{  \frac{2}{ m_0 \Omega}}
e^{ - \gamma t  - \mu \sin \nu t }  \beta_R, ~~
\Delta_p = \sqrt{ 2 m_0 \Omega } e^{ \gamma t +  \mu \sin \nu t }
\beta_I .
\end{equation}
Here $\beta (t)$ satisfies
\begin{equation}
\frac{d}{dt} \beta (t) + i \Omega \beta(t) = - i 
\sqrt{ \frac{1}{ 2 \Omega M(t) } } F(t),
\end{equation}
and the solution is given by (\ref{beta}). 
As in the case of {\it Example B} this system cannot possess a CIS
because of the damping effect.

Before closing this section we find the variances in $p$ and $q$
for a quantum state $\left| n, t \right>$. 
From (\ref{acop}) and (\ref{Bb}) we find
\begin{eqnarray}
\langle n | [ \Delta q(t)]^2 | n\rangle &=&
  (2 n + 1) \frac{g_-(t)}{2 \omega_I} ,
\\
\langle n | [ \Delta p(t)]^2 | n\rangle &=&
  (2 n + 1) \frac{\omega_I}{2 g_-(t) } 
  \left[   1 + \frac{g_0^2(t)}{\omega_I^2}    \right].
\end{eqnarray}
We note that as in Ref.~\cite{hckim} the linear driving terms don't change 
the width of a wave packet but change only the position
of the wave packet.

\section{discussion}

For the most general quadratic harmonic oscillator, 
we obtained the generalized invariant $I_T$ and its eigenstates. 
For these eigenstates, we have obtained the Schr\"{o}dinger wave function 
in a closed form, which can be applied to finding the
wave functions of our previous work~\cite{hckim}. 
The particular point is that the quantum wave function can be
written as the solutions of classical equation of motion with no linear driving terms. 
The linear driving terms [$F(t)$ and $G(t)$] merely shift the wave packets
of the eigenstates of the invariant {\it with no linear driving terms}
by the amount of $\Delta_q$ -- see (\ref{WF}). 
We have also investigated on the existence of a CIS,
and have found the conditions for the existence of CIS. 
For the CISs, the explicit form of Berry phases were obtained. 

The Hamiltonian $H_T(t)$ studied in this paper, by writing 
\begin{equation}
q = \left( \frac{1}{2 M \omega} \right)^{1/2}  ( a^\dagger + a),~
p = i \left(  \frac{M \omega}{2} \right)^{1/2} ( a^\dagger - a) ,
\end{equation}
can be written as 
\begin{equation}
H_T (t)  =  s_1 (t) a^2 +
            s_1^* (t) {a^\dagger }^2 +
            s_2 (t) [ a^\dagger a + 
             a a^\dagger ] + 
            s_3(t) a^\dagger + s_3^* (t) a,
            \label{last}
\end{equation}
where 
\begin{eqnarray}
s_1 = - \frac{i}{2}Y,~ 
s_2 =   \frac{ \omega}{2},~
s_3 = \frac{1}{\sqrt{2 m \omega}} \left( - F + i  m \omega G \right).
\end{eqnarray}
This Hamiltonian is studied in Ref.~\cite{yuen} to describe the photon 
process driven by a classical time-dependent source. 
The time-evolution operator
and geometric phase of this system appear in Ref.~\cite{XuQG91}. Therein the
physical quantities are expressed in terms of some auxiliary functions. While
in our formulation they are expressed in terms of classical solutions of
the corresponding GHO. 
Therefore we expect that our formulation is useful in describing 
the quantum optical properties of the model. 

\section*{Acknowledgement}

This work was partially supported by Korea Science and Engineering 
Foundation (KOSEF) and Non-Directed Research Fund, 
Korea Research Foundation, 1996. One of us (J.Y.J.) is supported
by Ministry of Education for the post-doctorial fellowship.


\begin{references}
\bibitem{Fulling}S. A. Fulling,  Aspects of Quantum Fields in Curved Space 
(Cambridge University Press, Cambridge, 1982).
\bibitem{Louisell73} W. H. Louisell,  Quantum Statistical Properties
of Radiation (John Wiley \& Sons, New York, 1973).
\bibitem{MeystreS91} P. Meystre and M. Sargent III,
 Elements of Quantum Physics (Springer-Verlag, New York, 1991).
\bibitem{lewis} H. R. Lewis Jr., Phys. Rev. Lett. {\bf 27}, 510 (1967); 
J. Math. Phys. {\bf 9}, 1976 (1968). 
\bibitem{LR} H. R. Lewis Jr. and  W. B. Riesenfeld,
J. Math. Phys. {\bf 10}, 1458 (1969). 
\bibitem{spkim}S. P. Kim, J. Phys. {\bf A 27}, 3927 (1994).
\bibitem{yeon}K. H. Yeon, H. J. Kim, C. I. Um, T. F. George and
L. N. Pandey, Phys. Rev.  A {\bf 50}, 1035  (1994).
\bibitem{jyji2}J. Y. Ji, J. K. Kim, S. P. Kim and K. S. Soh,
Phys. Rev. A {\bf 52}, 3352 (1995).
\bibitem{lawande}D. C. Khandekar and S. V. Lawande, J. Math. Phys.
{\bf 20}, 1870 (1979).
\bibitem{lo}C. P. Lo, Phys. Rev. {\bf A 43}, 404 (1991).
\bibitem{hckim}H. C. Kim, M. H. Lee, J. Y. Ji and J. K. Kim, 
Phys. Rev. A {\bf 53}, 3767 (1996).
\bibitem{gao}X.-C. Gao, J.-B. Xu and T.-Z. Qian, 
Phys. Rev A {\bf 44}, 7016 (1991).
\bibitem{jyji}J. Y. Ji, J. K. Kim and S. P. Kim,
Phys. Rev. A {\bf 51}, 4268  (1995).
\bibitem{anandan}Y. Aharonov and J. Anandan, 
Phys. Rev Lett. {\bf 58}, 1593 (1987).
\bibitem{cal} P. Caldirola, Nuovo Cimento, {\bf 18}, 393 (1941).
\bibitem{kanai} E. Kanai, Prog. Theor. Phys {\bf 3}, 440  (1948). 
\bibitem{report} D. J. Moore, Phys. Rep. {\bf 210}, 1 (1991).
\bibitem{yuen}H. P. Yuen, Phys. Rev. {\bf A 13}, 2226 (1976).
\bibitem{XuQG91} J.-B. Xu, T.-Z. Qian and X.-C. Gao, 
Phys. Rev. A {\bf 44}, 1485 (1991).
\end{references}
\end{document}